\documentclass[aps]{revtex4}
\usepackage{graphicx}

\begin{document}

\title{Scattering length determination from trapped pairs of atoms}
\author{Sanjiv Shresta}\email{sanjiv.shresta@nist.gov}
\affiliation{Atomic Physics Division, NIST, Gaithersburg, MD}
\author{Eite Tiesinga}%\email{eite.tiesinga@nist.gov}
\affiliation{Atomic Physics Division, NIST, Gaithersburg, MD}
\author{Carl Williams}%\email{carl.williams@nist.gov}
\affiliation{Atomic Physics Division, NIST, Gaithersburg, MD}
\date{\today}

\begin{abstract}
A method is described for estimating effective scattering lengths via spectroscopy on a trapped pair of atoms. The method relies on the phenomena that the energy levels of two atoms in a harmonic trap are shifted by their collisional interaction. The amount of shift depends on the strength of the interaction (i.e. scattering length).
By combining the spectra of the trap state energy levels and a suitable model for the effective scattering length, an estimate for the latter may be inferred. Two practical methods for measuring the trap spectra are proposed and illustrated in examples. The accuracy of the scheme is analyzed and requirements on measurement precision are given.
%This method may provide a high precision means of determining scattering lengths for known and unknown species, with applications to energy dependent collisional shifts in clock experiments.
\end{abstract}
\maketitle

\section{Introduction}
An important parameter in the field of atomic collisions, which can characterise much of a particle's collision properties, is the scattering length. Defined in terms of the s-wave phase shift at zero collision energy, validity of the scattering length is restricted to the Wigner threshold regime. At higher energies the scattering length description begins to break down, and the total cross-section deviates from $4\pi$ times the scattering length squared. Despite that fact, the concept of a scattering length remains useful when widened
%to include energy dependence, which is acheived
by simply defining an \emph{effective} scattering length at any collision energy, still in the s-wave approximation~\cite{Block02,Bolda02} or for higher partial waves~\cite{Stock04}. 

For cesium in particular the collision energy dependence of the scattering length takes on special importance. Existing methods for estimating scattering lengths have been very successful with rubidium and lighter alkali atoms~(see Refs.~\cite{Heinzen99,PethickSmith} for a thorough review and the following references for more recent work utilizing Bose-Einstein condensate~(BEC) mean field expansion~\cite{SantosTannoudji01,VolzRempe03}, rethermalization~\cite{FerrariSimoni02,SchmidtSimoni03}, ion flux rates~\cite{SeidelinWestbrook04}, and Bragg spectroscopy~\cite{TheisDenschlag04}). However, measurements of cesium scattering lengths show an unusual degree of disagreement. For example, the triplet scattering length of $\mbox{}^{133}$Cs has been predicted to be many various values ranging from -1100$a_o$ to 2400$a_o$ (with $a_o$ being a Bohr radius). The source of this disagreement is that the $\mbox{}^{133}$Cs triplet scattering length exhibits strong variation with collision energy, due to a nearly bound above threshold resonance~\cite{VerhaarChu93,ArndtDalibard97,LegereGibble98,KokkelmansGibble98,HopkinsFoot00,DragPillet00}. In different collision energy regimes all the quoted values can be correct. Most recently, the $\mbox{}^{133}$Cs triplet scattering length at near zero collision energy has been predicted to be 2440(25)$a_o$ by first fitting predictions of the expected interatomic Hamiltonian to the most accurate Feshbach resonance measurements available~\cite{LeoJulienne00,ChinWilliams04}. 

Although the energy dependence of the Cs triplet scattering length has been directly observed using thermal equilibration~\cite{ArndtDalibard97,HopkinsFoot00} and ballistic scattering~\cite{LegereGibble98}, since those techniques rely on thermal clouds of atoms, they are limited to thermal spreads in collision energy. Their estimates will be averages over a spread of collision energies, with the smallest such spread being $\Delta E/h=30$~kHz in Ref.~\cite{LegereGibble98}. In regimes of strong variation with collision energy, such an effect could be considerable. 

Described here is a scheme for determining the effective scattering length over a range of collision energies which does not suffer from that drawback. It exploits a well known effect on trapped pairs of atoms: the interaction between the atoms will alter the trap vibrational eigenenergies. The main idea of the scheme is to infer the scattering length by measuring the transition energies between vibrational eigenstates. Both the magnitude and the sign of the effective scattering length can be inferred and the range of collision energies at which the scattering occurs is controlled by choosing the trap frequency. The tradeoff for increased precision of collision energy is increased sensitivity to measurement uncertainty and noise. Error analysis shows that successful application of this method~(e.g.~approx~$10\%$ accuracy in the scattering length estimate) will require spectroscopic measurement uncertainty and trap frequency stabilisation of $10^{-3}$ and $10^{-4}$ of the trap frequency, respectively. 

Finally, an estimate of the energy dependence of the effective scattering length will be a useful tool. For example, in cesium fountain clocks, collision energy dependent frequency shifts are currently one of the main limiting factors in the precision of the clocks. With a prediction of the effective scattering length, the clock shifts could be estimated from a known expression based on the Boltzmann equation~\cite{Tiesinga92,GibbleChu93,VerhaarChu93,LeoWilliams01}. Another example is the predicted effect, from energy dependence in the scattering length, on the ground state energy of a tightly confined BEC~\cite{FuWangGao03}.

The organization of the paper is as follows. In Section~II a brief exposition of the theoretical background is given and a few important equations are introduced. In Section~III a procedure for estimating the effective scattering length from knowledge of trap spectra for a pair of trapped atoms is outlined. In Section~IV two proposed measurement schemes are described and examples of scattering length extraction for each are shown. A discussion of the schemes including error estimates is then made, followed by, in Section~V, some concluding remarks.

\section{Background}
The proposed scheme for estimating elastic scattering lengths from trap spectroscopy is based on the combination of two theoretical ideas. The first is the collisional energy shift of trap vibrational levels as computed in Refs.~\cite{Busch97,Blume02,Block02,Bolda02}. The second is a resonance model for the effective scattering length in the presence of open channel (virtual) and closed channel (Feshbach) resonances as summarized in Ref.~\cite{MarcelisKokkelmans04}.

In the first theoretical ingredient, the interaction between two atoms, each of mass $m$, is modelled with an s-wave delta function pseudopotential which takes the form
\begin{equation}\label{pseudopotential interaction}
V = \frac{4\pi a\hbar^2}{m} \delta({\bf r}) \frac{\partial}{\partial r} r.
\end{equation}
The strength of the interaction is parameterized by the scattering length, $a$, and involves only the relative coordinate, $r$. Thus, in a harmonic trap, for which the center of mass motion can be separated from the vibrational motion, the center of mass eigenenergies are of the usual harmonic oscillator states.
The relative coordinate eigenstates are altered by the interaction so that in a spherically symmetric harmonic trap the eigenenergies are the solutions of the equation
\begin{equation}\label{Busch3d}
\frac{a}{l}= \frac{1}{2} \tan{\bigg(\frac{\pi E}{2} +\frac{\pi}{4}\bigg)} \mbox{ }\frac{\Gamma(1/4+E/2)}{\Gamma(3/4+E/2)}
\end{equation}
with $E$ in units of $\hbar\omega$, $l=\sqrt{2\hbar/m\omega}$ being the characteristic size of the trap for the relative motion, and $a$ being the scattering length. This expression is sufficient to make accurate estimates of trap vibrational eigenenergies as long as the trap is shallow enough that the collision energy of the atoms is in the Wigner threshold regime and the condition $a/l << 1$ holds~\cite{Tiesinga00,Blume02,Block02}.

In order to estimate trap eigenenergies at higher trap frequencies, Eq.~(\ref{Busch3d}) was generalized to include energy dependence in the pseudopotential strength~\cite{Block02,Bolda02}. The new condition is the same as Eq.~(\ref{Busch3d}) with an energy dependent scattering length,
\begin{equation}\label{Bolda3d}
\frac{a(E)}{l}= \frac{1}{2} \tan{\bigg(\frac{\pi E}{2} +\frac{\pi}{4}\bigg)} \mbox{ }\frac{\Gamma(1/4+E/2)}{\Gamma(3/4+E/2)} \equiv f(E),
\end{equation}
also with $E$ in units of $\hbar\omega$. It was shown in Ref.~\cite{Bolda02} that Eq.~(\ref{Bolda3d}) accurately predicts the eigenenergies for tight traps. The new condition of validity is that the trap characteristic length be much longer than the van~der~Waals potential scale length~(see appendix of Ref.~\cite{WilliamsGould99}), rather than the scattering length. The right hand side of Eq.~(\ref{Bolda3d}) is for convenience defined as the intercept function, $f(E)$.
The function $a(E)$ is an effective scattering length defined in terms of the s-wave phase shift, $\delta_o(k)$,
\begin{equation}\label{effective scattering length}
a(E) = -\frac{\tan \delta_o(k)}{k},
\end{equation}
which incorporates the combined effects of all molecular interactions at energy $E=\hbar^2 k^2/m$.  The intersections of the intercept function and the effective scattering length are the eigenenergies of the relative coordinate motion.

The second ingredient is a functional form for the energy dependence of the scattering length, Eq.~(\ref{effective scattering length}). A reasonably accurate model, which includes the effect of a single resonance is
\begin{equation}\label{general model fit}
a(E) = a_b +\frac{\alpha}{E-E_{r}},
\end{equation}
with $a_b$ being a constant background piece, and $\alpha$ and $E_r$ defining the energy dependence~\footnote{A more exact form based on partial resummation of the Lippman-Schwinger equation has recently been derived~\cite{MarcelisKokkelmans04}. The form used here approximates that expression, but is sufficient for our purpose.}. This form for Eq.~(\ref{general model fit}) is a generic simplification of the correct resummed expression and only approximates the true resonance behavior, but for our purposes, which is fitting away from the resonance, it is sufficient. Comparison of this form with numerical data from a close coupling code based on the most accurate Cs potentials available~\cite{ChinWilliams04} shows excellent agreement. Fig.~(1) shows numerically calculated scattering lengths for a collision between doubly-polarized cesium atoms and a fit to the simulated data with Eq.~(\ref{general model fit}). The inset of Fig.~(1) shows a magnified region with the intercept function superimposed. In this case a single resonance fit is sufficient, however, a dual resonance model may be necessary in some cases, to account for closely spaced resonances.
\begin{figure}
\begin{center}
\includegraphics[width=8cm]{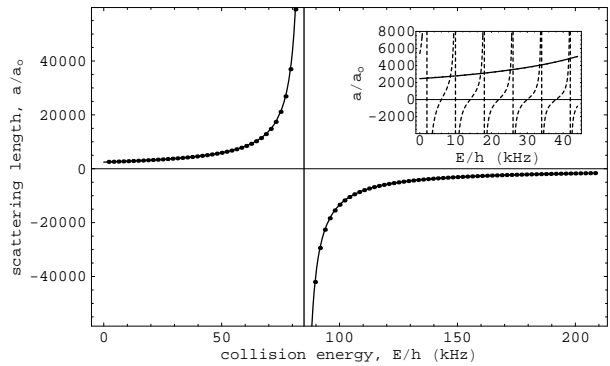}
\caption{Numerically computed data (dots) for the scattering length of a pair of doubly
polarized cesium atoms versus collision energy compared to a best
fit (solid line) of the form of Eq.~(\ref{general model fit}). The
inset shows a close-up with the intercept function superimposed as
a dashed line. The intersections of the effective scattering
length and the intercept function are the vibrational
eigenenergies. ($a_o =0.0529\mbox{ nm}$ is the Bohr radius.)}\label{diagscat}
\end{center}
\end{figure}

\section{Inversion Procedure}
As mentioned in the previous section, the eigenenergies of the relative motion are found at the intersections of the effective scattering length, $a(E)$, and the intercept function, $f(E)$. Conversely, if the eigenenergies are known, the effective scattering length can be extracted by matching its intersections with the intercept function to the known eigenenergies. Procedurely, that can be accomplished by solving a set of equations consisting of,
\begin{equation}\label{maineq}
a(E_i) = l f(E_i),
\end{equation}
for discrete values of the collision energy, $E_i$. The unknowns are the parameters of the model for $a(E)$ and any single absolute eigenenergy of the trap. The last of those unknowns, an absolute eigenenergy~(relative to trap zero), is included as a fitting parameter because in any experiment only energy differences are observed. Thus for a single resonance model there are four unknowns ($a_b$, $\alpha$, $E_r$ and e.g. $E_0$) and at least four equations for any four eigenenergies and measurements of the energy differences between those eigenenergies are needed.

In practice the equations are non-linear and do not allow a simple solution. Instead we use a two-step process to determine the unknown parameters. For example, for the single resonance model, choosing the lowest four eigenenergies, we do a least squares fit of the parameters $a_b$, $\alpha$, and $E_r$ to the equation $a(E_i) - l f(E_i)$ for a fixed value of $E_0$. This will yield a least squares error for each best fit, denoted by
\begin{equation}\label{chisq}
\chi^2(E_0) = \sum_i | a(E_i) - l f(E_i) |^2,
\end{equation}
where the summation is over the remaining energies.
A near zero minimum of $\chi^2(E_0)$ will locate a reference eigenenergy $E_0$ which gives a good fit.
If there is more than one near zero minimum in the function $\chi^2 (E_0)$, either more transition energies or outside estimates of the scattering length can be used to decide which is the true eigenenergy. Moreover, it should be remembered that the intercept function has asymptotes related to the `tangent' function contained in it. The intersection points of the effective scattering length and the intercept function must be within those asymptotes and there can be only one eigenenergy per interval.

\section{Measurement Examples}
There are two issues that must be addressed in order to make this method experimentally viable. The first is that atoms must be trapped in isolated pairs. The second is that the transition energies between trap vibrational eigenstates must be extracted. Both of these issues will depend on the particular atoms being measured. In the examples described in the following subsections, pairs of hyperfine ground state Cs atoms with no relative angular momentum, assuming a spherical and harmonic trapping potential, will be analyzed.

In regards to the first issue, techniques currently exist for creating Mott-insulator states in optical lattices with one or two atoms per lattice site from a single component BEC~\cite{GreinerBloch02}. For measurements of the scattering length between identical atoms, such techniques will properly set up the system. Pairs of homonuclear atoms in different hyperfine states or magnetic sublevels could also be initialized if the atoms may be individually addressed in a spin dependent lattice~\cite{DeutschJessen98,MandelBloch03}. In the case of different atoms, techniques have recently been proposed~\cite{DamskiZoller03,ModugnoInguscio03}.

In regards to the second issue, there are two ways in which the energy level spacings can be measured. Which one to use depends on the collisional lifetime of the hyperfine state and magnetic sublevel of the pair of atoms. The distinction between stable and unstable states is important because pairs of trapped atoms with high atomic densities~($\sim 10^{14}\mbox{ cm}^{-3}$), and long measurement pulses~($\sim 100\mbox{ ms}$) will be necessary in order to measure the collisional shift of vibrational eigenenergies. If the atomic pair's states are stable with respect to transitions to other channels on a timescale longer than the measurement time, then Raman spectroscopy techniques can be used to measure the vibrational transition frequencies. If, on the other hand, the pair's states will scatter to other channels on times shorter than the measurement time, another method based on measuring Rabi oscillations, can be applied. Both methods are illustrated through examples in the following subsections. In each case scattering lengths from the transition amplitudes obtained via close coupling numerical computations~\cite{ChinWilliams04} are used to generate simulated spectral data~(e.g.~Figs.~(2)~and~(5)). The simulated spectra are then put through the procedure of Section~III to re-infer the scattering lengths.

\subsection{Raman Spectroscopy}
The case of a pair of Cs atoms in their doubly-polarized internal state, $| f_1=4,m_1=4 \rangle |f_2=4,m_2=4 \rangle$, falls into the category of a stable hyperfine channel, where $f_\alpha$ and $m_\alpha$ denote the internal hyperfine state and the magnetic sublevel of atom $\alpha$. The presence of a small external magnetic field such that magnetic splittings are large compared to the trapping frequencies is assumed. The relative coordinate vibrational energy levels for such a pair of atoms in a harmonic trap with a frequency of 4~kHz is shown in Fig.~(2).
\begin{figure}
\begin{center}
\includegraphics[width=8cm]{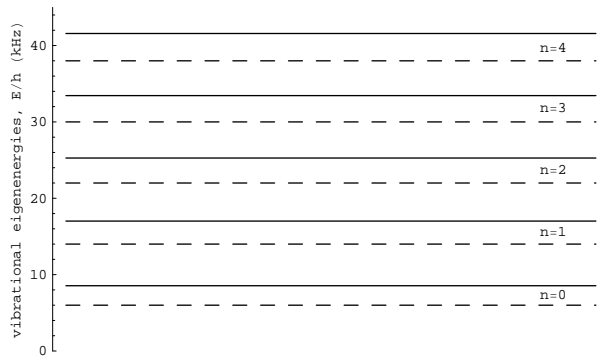}
\caption{Energy level diagram for a doubly polarized pair of Cs atoms in a 4~kHz harmonic trap. The ground ($n=0$) through 4th excited relative coordinate vibrational eigenenergies are shown. The dashed lines are the non-interacting eigenenergies and the solid lines are the eigenenergies including the pseudopotential interaction of Eq.~(\ref{pseudopotential interaction}).
}\label{levels}
\end{center}
\end{figure}
The dotted lines are the non-interacting relative coordinate vibrational eigenenergies and the solid lines are the eigenenergies including a pseudopotential interaction. The spacing between the eigenenergies is roughly 8~kHz, as expected for symmetric bosonic eigenfunctions. The interacting eigenlevels are shifted to higher energies relative to the non-interacting ones because the scattering length is positive in the shown energy range (see Fig.~(1)), making the interaction repulsive.

In this case the vibrational energy spacings can be measured with Raman spectroscopy by setting the virtual intermediate state far off-resonant from an excited molecular state.
Experiments in which vibrational transition frequencies have been measured exist in the literature. For example, in Ref.~\cite{LounisGrynberg93} the authors use four-wave mixing and detection of a phase-conjugate signal~(see Ref.~\cite{GrynbergRobillard01} for a review) to measure the quantization of atomic motion in an lattice. By such a process they are able to see vibrational transitions between adjacent levels. Recently, in Ref.~\cite{RomBloch04}, the authors have been able to resolve vibrational transitions using photo-association spectrocopy. In this latter reference the experimental setup is an optical lattice filled with isolated pairs of Rb in their doubly-polarized hyperfine state. Their configuration is exactly what is necessary in the present scheme. 
In both of those implementations the resolution of the transition frequencies were insufficient for purposes here. A combination of the ideas and techniques of both experiments might be used for gaining the needed accuracy. 

Assuming that the energy level spacings are perfectly measured for the lowest few vibrational eigenstates,
the least squares error, $\chi^2(E_0)$, is shown in the inset of Fig.~(3) versus the ground state energy within the range of possible values.
\begin{figure}
\begin{center}
\includegraphics[width=8cm]{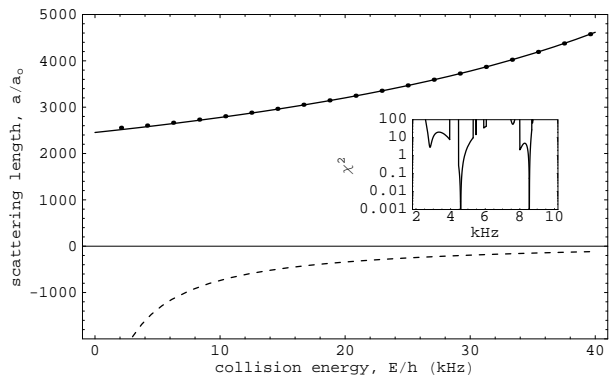}
\caption{ The scattering length analysis for a pair of doubly polarized atoms in a $4$~kHz trap is illustrated. The inset shows the least squares error, $\chi^2(E_0)$, of best fits to the simulated spectra versus ground state energy. There are two possible fits at $E_0/h=8.549\mbox{ kHz}$ and $E_0/h=4.580\mbox{ kHz}$. The dots are the original numerically generated scattering length data, the solid line is the scattering length estimate for $E_0/h=8.549\mbox{ kHz}$, and the dashed line is the scattering length estimate for $E_0/h=4.580\mbox{ kHz}$.
}\label{chiandscat}
\end{center}
\end{figure}
Inspection of this inset shows that there are two good fits for the ground state energy, at $E_0/h=4.580\mbox{ kHz}$ and $E_0/h=8.549\mbox{ kHz}$. The scattering length estimate obtained using $E_0/h=4.580\mbox{ kHz}$, the scattering length estimate for $E_0/h=8.549\mbox{ kHz}$, and the original numerically generated data are shown together in Fig.~(3) within the energy range of the ground to 3rd excited states. Agreement with the known approximate value for the scattering length clearly rules out the $E_0/h=4.580\mbox{ kHz}$ fit and choses $E_0/h=8.549\mbox{ kHz}$ as the correct reference energy. If a best value for the reference ground state energy can not be clearly identified, then more data will be required~(e.g. spectra at another trap frequency). The fit parameters for $E_0/h=8.549\mbox{ kHz}$ are $a_b/a_o= 36$, $\alpha/a_o = -2.049\mbox{ }10^5\mbox{ kHz}$ and $E_r/h = 84.72\mbox{ kHz}$ (with $a_o =0.0529\mbox{ nm}$ being the Bohr radius).

The accuracy of the scattering length estimation in Fig.~(3) is nearly perfect. In the case where there is uncertainty in the measurements, that will not be the case. Figure~(4) shows a plot of the accuracy of the scattering length estimate versus the uncertainty in measurements of the line positions.
\begin{figure}
\begin{center}
\includegraphics[width=8cm]{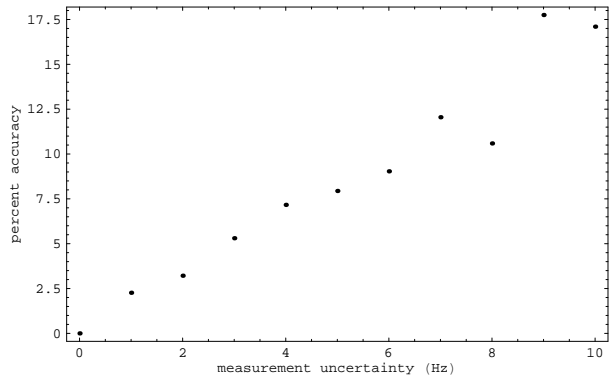}
\caption{The percent deviation between the scattering length estimated from simulated data with random measurement uncertainty applied and the exact scattering length estimations. Each point is averaged over the ground to 3rd excited state energy range and is the mean of twenty simulations.}
\end{center}
\end{figure}
The points are the integrated average percent difference
\begin{equation}\label{error}
\bigg\langle  \frac{1}{E_3 -E_0} \int_{E_0}^{E_3} dE \frac{|a_{err}(E) -a(E)|}{|a(E)|} \bigg\rangle
\end{equation}
over the energy range between the ground to 3rd excited state. The scattering length, $a_{err}(E)$, is obtained from simulated measurements with uniformly distributed random noise, whereas the scattering length, $a(E)$, is obtained with no measurement noise. The $\langle \cdots \rangle$ indicate an average over simulations. The points in Fig.~(4) are scattered because each point is an average of only twenty simulations.
The accuracy varies from approximately $2\%$ for $1\mbox{ Hz}$ uncertainty to $17\%$ for $10\mbox{ Hz}$ uncertainty. In the case of a $4\mbox{ kHz}$ trap, that level of sensitivity corresponds to a required measurement precision of one in one thousandth ($10^{-3}$) of a trap frequency for a $10\%$ accuracy in the scattering length estimation.

\subsection{Rabi Oscillation Beats}
For a pair of trapped Cs atoms in the same hyperfine state and with magnetic sublevels such that the total projection quantum number, $M=m_1+m_2$, is non-extremal, there will be a degenerate set of magnetic sublevel pairs. If collisions between the atoms couple the magnetic sublevels but do not change $M$, as is the case for alkali-metal atoms, such states will fall into the category of states that elastically scatter into other channels. For example, $\mbox{Cs}_2$ in the clock state $| f_1=3,m_1=0 \rangle |f_2=3,m_2=0 \rangle$, will couple through collisions to others channels that have $M=0$. Such transitions will not cause trap loss and the pair of atoms will coherently oscillate between the coupled channels. In reality the channels are only nearly degenerate, with energy differences governed by the quadratic Zeeman splitting, which must be less than the trapping frequency.

The molecular states $|f_1=3,f_2=3,F,M\rangle$ with total spin, $\vec{F}=\vec{f}_1 +\vec{f}_2$, and projection, $M$, quantum numbers are stable under s-wave collisions. Consequently, each molecular state has its own effective scattering length. The energy level diagram for different $F$'s in a 4~kHz harmonic trap including the s-wave interaction is shown in Fig.~(5). In the present example all scattering lengths are negative, in contrast to the example of the previous section. Note that Bose symmetry of the atoms restricts the pair's states to even $F$'s.
\begin{figure}\label{ddCslevel}
\begin{center}
\includegraphics[width=8cm]{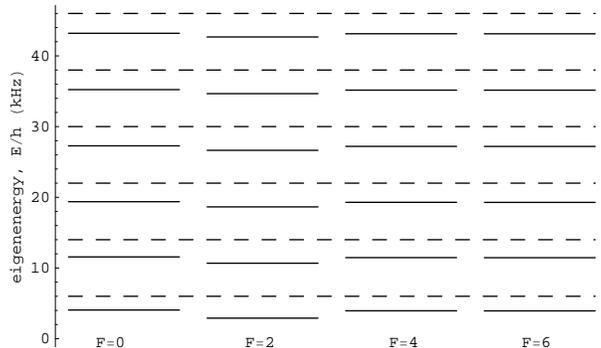}
\caption{Energy level diagram for a pair of $f_1=f_2=3$ hyperfine Cs atoms in a 4~kHz harmonic trap. The dashed lines are the free vibrational eigenenergies. The solid lines are the eigenenergies for, from left to right, $F=$0,~2,~4, and~6 and $M=0$ in the molecular basis.}
\end{center}
\end{figure}

Coherent oscillations can be exploited to gain information about the vibrational energy levels in the following way. First load a lattice in the lin$\perp$lin laser configuration with $| f_1=3, m_1=3 \rangle$ and $| f_2=3, m_2=-3 \rangle$ cesium atoms singly loaded in the resulting intertwined lattices, respectively. Then change the laser configuration to lin$\parallel$lin so that pairs of atoms are trapped in the same optical potential and do so quickly enough that many vibrational states are excited. 
In this way the pair of atoms will have been prepared in a non-eigen internal and external state. Its state will then Rabi oscillate in time through a superposition of many eigenfrequencies. A simulation of the population of $|f=3, m=0 \rangle$ versus time is shown in Fig.~(6). Other populations show a similar behavior.
\begin{figure}\label{ddCsosc}
\begin{center}
\includegraphics[width=8cm]{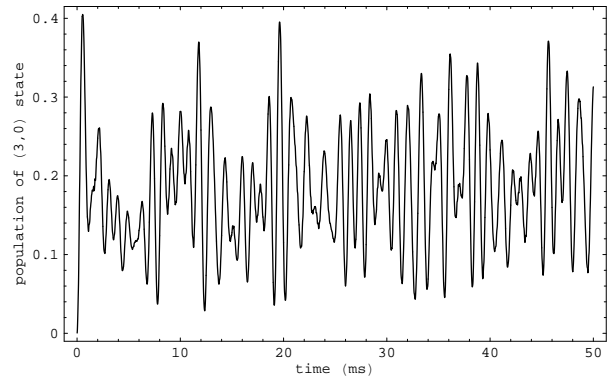}
\caption{Oscillation of the population of $| f=3, m=0 \rangle$ states given an initial $| f=3, m=3 \rangle | f=3, m=-3 \rangle$ internal state and a spatial wavefunction that is not a vibrational eigenstate of the trap. The trap frequency is 4 kHz.}
\end{center}
\end{figure}

A similar concept was demonstrated, in a $\mbox{Rb}$ BEC experiment in which the coupling between different internal states was mediated by applied laser fields~\cite{MatthewsCornell99,WilliamsHolland00}. In that experiment the vibrational eigenstates for the two internal states differed because each internal state felt a different trapping potential. The result of observing the population of one of the states was Rabi oscillation beats, attributed to the coupling between the internal and external state dynamics. Coherence times of hundreds of milliseconds was observed.

The differences in vibrational frequencies, which are the quantities needed to estimate the scattering length, can be obtained by a Fourier transform of the oscillation of the $| f=3,m=0 \rangle$ population. As shown in Fig.~(7) the Fourier transform of the Rabi oscillation behavior has peaks at all the transition frequencies of the system. Closer inspection of Fig.~(7) shows separate groups of spectral components. From left to right they correspond to transitions involving changes in the vibrational quantum number of $\Delta$n=0,~1,~2,~3 and so on. 
\begin{figure}\label{ddCsfour}
\begin{center}
\includegraphics[width=8cm]{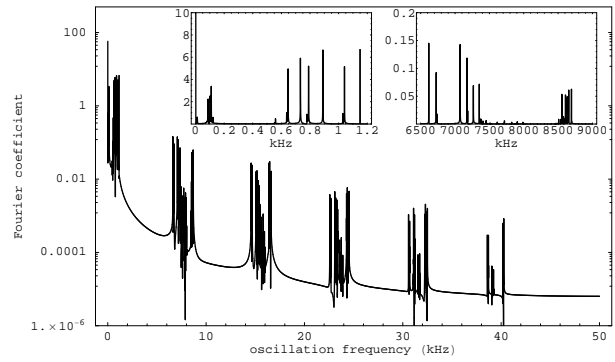}
\caption{Fourier transform of the population in Fig.~(6). The left inset is a zoom on the $\Delta$n=0 Fourier components and the right inset is a zoom of the $\Delta$n=1 Fourier components.}
\end{center}
\end{figure}
\clearpage
The left inset of Fig.~(7) is the fine structure on the $\Delta$n=0 spectral components and the right inset is the fine structure on the $\Delta$n=1 set of components. Since Fig.~(7) is based on Fig.~(5), the fine structure lines can straightforwardly be assigned to their respective eigenfrequency differences. 

In reality an energy level diagram such as in Fig.~(5) is unavailable a priori. However, with reasonable estimates for the scattering lengths the structure of the frequencies can be predicted and will be close to the true eigenenergy structure. With the simulation as a guide, the lines can be identified with their transitions. The assignment can then be used to find, for example, the lowest three transition frequencies for each $F$. A sample of the results of a least squares fit for $F=0$, performed as in the previous section, is shown in the inset of Fig.~(8). Inspection of this inset shows that, as before, there are two good fits for the ground state energy, at $E_0/h=4.038\mbox{ kHz}$ and $E_0/h=8.022\mbox{ kHz}$. The scattering length estimate obtained using $E_0/h=4.038\mbox{ kHz}$, the scattering length estimate for $E_0/h=8.022\mbox{ kHz}$, and the original numerically generated data are shown together in Fig.~(8). As before, agreement with the known approximate value for the scattering length clearly rules out one fit and choses $E_0/h=4.038\mbox{ kHz}$ as the correct reference energy.

The agreement in this idealized case is better than $1\%$.
\begin{figure}\label{rabiscat0}
\begin{center}
\includegraphics[width=8cm]{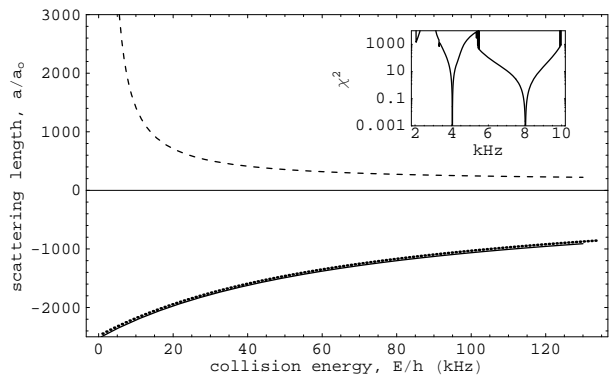}
\caption{Scattering length estimates for $F=0$ in a 4~kHz trap are shown. The dashed line shows the scattering length estimation for $E_0/h=8.022\mbox{ kHz}$, the solid line shows the scattering length estimation for $E_0/h=4.038\mbox{ kHz}$, and the dotted line is the original numerically generated scattering length data. The upper inset shows the best fit $\chi^2$-error versus ground state energy, which exhibits two, rather than just one, possible fit values for the reference ground state energy.}
\end{center}
\end{figure}
An estimate of the percent accuracy computed from simulated measurements with random error included, as in Eq.~(\ref{error}), is shown in Fig.~(9).
\begin{figure}\label{rabierror}
\begin{center}
\includegraphics[width=8cm]{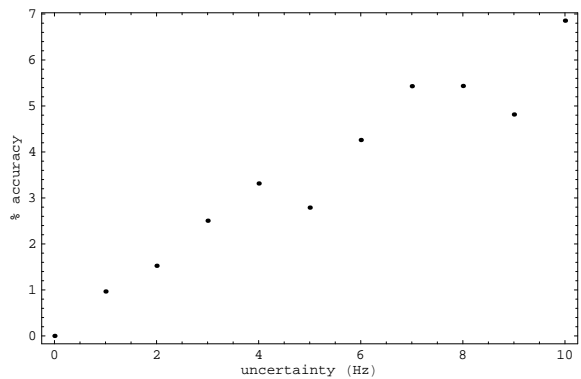}
\caption{The percent deviation between exact scattering length estimations and the scattering length estimated from simulated data with random measurement uncertainty applied. Each point is averaged over the ground to 3rd excited state energy range and is the mean of twenty simulations.}
\end{center}
\end{figure}
The dependence of the error in the predictions versus the measurement uncertainty is qualitatively the same as for the Raman technique of Fig.~(4). The accuracy varies from approximately $1\%$ for $1\mbox{ Hz}$ uncertainty to $7\%$ for $10\mbox{ Hz}$ uncertainty. As in the previous section, for a $4\mbox{ kHz}$ trap that level of sensitivity requires a measurement precision of on the order of one in one thousandth ($10^{-3}$) of a trap frequency for a $10\%$ accuracy in the scattering length estimate.

\begin{figure}
\begin{center}
\includegraphics[width=8cm]{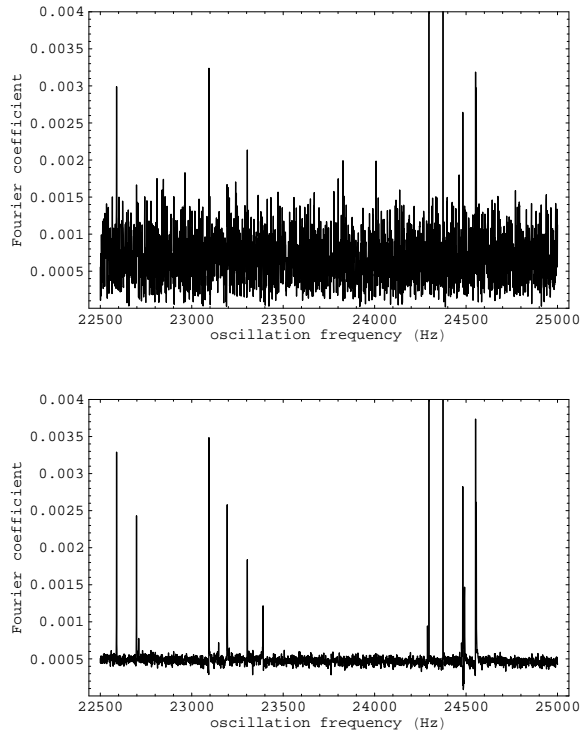}
\caption{These figures illustrate the effect of trap frequency jitter on the readability of the Fourier peaks. The top plot is the $\Delta n=3$ fine structure when trap frequency jitter of $10^{-3}$ is included, and the bottom figure is the same for frequency jitter of $10^{-4}$.}\label{rabinoise}
\end{center}
\end{figure}

In addition to measurement error in the oscillation frequencies, this method is susceptible to drift of the Rabi oscillations due to trap frequency instability. Such noise, as would be caused by laser intensity fluctuations, could cause the noise level to rise above the Fourier peaks of Fig.~(7). An illustration of this is shown in the top and bottom plots of Fig.~(\ref{rabinoise}), in which the Fourier peaks are simulated with $10^{-3}$ and $10^{-4}$ trap frequency jitter included, respectively. In order to have adequate readibility thus requires on the order of $10^{-4}$ frequency stabilization.

\subsection{Discussion}
The previous two simulated examples have assumed, in the ideal case, noiseless measurements with less than $1\mbox{ Hz}$ uncertainty. The resulting agreement with the original numerical estimates was then nearly exact. In an experimental application there will be noise which places constraints on measurement precision and thus the spectral line resolution. That issue has been addressed in Figs.~(4)~and~(9), in which it is shown that a measurement uncertainty of $10^{-3}$ of the trap frequency will give on the order of $10\%$ accuracy for the scattering length. In the schemes described above, that degree of resolution will in turn require a measurement time of a few hundred milliseconds, and commensurate supression of broadening due to anharmonicity, inhomogeneity, jitter in the lattice, and irreversible collisonal processes in the lattice. A second unwanted effect of noise is that it may overwhelm the signal. That effect, coming from jitter in the trap frequency, has been quantified for the Rabi oscillation scheme in Fig.~(\ref{rabinoise}). It is seen that the stabilization of the trap frequency must be approximately $10^{-4}$ in order for sufficient signal to be identifiable. Both of the minimum requirements -- on line resolution and stabilization of trap frequency -- are beyond current experimental capabilities. 
However, they are close enough to current capabilities that the requisite improvements are on the horizon.

\section{Concluding Summary}
In summary, a method has been described for estimating energy dependent effective scattering length from spectroscopic measurements of the vibrational transition energies of a pair of trapped atoms. Two examples were illustrated for the case in which the state of the pairs of atoms is stable against collisional transitions and that in which it is not stable. Experimental implementation of this method will require spectral resolution of better than $10^{-3}$ of the trap frequency and trap frequency stabilization of $10^{-4}$. Though beyond current experimental precision, as improvements are made in optical lattice technology, the scheme described here will be a useful method for finding precision estimates of scattering properties.

\end{document}